\documentclass[sigconf, nonacm]{acmart}
\usepackage{subfig}
\graphicspath{{samples/}}
\AtBeginDocument{%
  }

\setcopyright{none}
\copyrightyear{2026}
\acmYear{2026}
\acmDOI{XXXXXXX.XXXXXXX}
\acmConference[SIGIR'26]{Proceedings of the ACM Web Conference 2026}{July 20--24, 2026}{Melbourne, Australia}
\acmISBN{}

\begin{document}

\title{Multimodal Enhancement of Sequential Recommendation}

\author{Bucher Sahyouni}
\affiliation{%
  \institution{University of Surrey}
  \city{Guildford}
  \country{United Kingdom}
}
\email{bs00826@surrey.ac.uk}

\author{Matthew Vowels}
\affiliation{%
  \institution{The Sense, CHUV}
  \city{Lausanne}
  \country{Switzerland}
}
\affiliation{%
  \institution{Kivira Health}
  \city{New York}
  \state{NY}
  \country{USA}
}
\email{matthew.vowels@unil.ch}

\author{Liqun Chen}
\affiliation{%
  \institution{University of Surrey}
  \city{Guildford}
  \country{United Kingdom}
}
\email{liqun.chen@surrey.ac.uk}

\author{Simon Hadfield}
\affiliation{%
  \institution{University of Surrey}
  \city{Guildford}
  \country{United Kingdom}
}
\email{s.hadfield@surrey.ac.uk}

\renewcommand{\shortauthors}{Sahyouni et al.}

\begin{abstract}
  We propose a novel recommender framework, MuSTRec (Multimodal and Sequential Transformer-based Recommendation), that unifies multimodal and sequential recommendation paradigms. MuSTRec captures cross-item similarities and collaborative filtering signals, by building item-item graphs from extracted text and visual features. A frequency-based self-attention module additionally captures the short- and long-term user preferences. Across multiple Amazon datasets, MuSTRec demonstrates superior performance (up to 33.5\% improvement) over multimodal and sequential state-of-the-art baselines. Finally, we detail some interesting facets of this new recommendation paradigm. These include the need for a new data partitioning regime, and a demonstration of how integrating user embeddings into sequential recommendation leads to drastically increased short-term metrics (up to 200\% improvement) on smaller datasets. Our code is availabe at https://anonymous.4open.science/r/MuSTRec-D32B/ and will be made publicly available.
\end{abstract}

\ccsdesc[500]{Information systems~Recommender systems}
\ccsdesc[300]{Computing methodologies~Neural networks}
\ccsdesc[500]{Information systems~Multimedia information systems}

\keywords{Multimodal, Sequential, Recommender System, GNN, Transfromer}

\received{20 February 2007}
\received[revised]{12 March 2009}
\received[accepted]{5 June 2009}

\maketitle

\section{Introduction}
\label{sec:intro}
Recommender systems are the cornerstone of countless online services, providing personalised recommendations to users based on preferences inferred from their interactions. To improve this personalisation, modern recommender system have sought to extract more meaning from this interaction data. The two dominant contemporary paradigms are Sequential recommender systems\cite{hidasi2016} (capturing the temporal evolution of user behaviour) and multimodal recommender systems\cite{zhang2021ai} (integrating additional item data such as written descriptions, pictures or trailers to extract hidden similarities).
We propose MuSTRec which unifies both paradigms while also exploiting an interaction graph to share training constraints between similar users and items.

The main advantage we gain from explicitly considering sequential recommendation in our approach is the ability to encode temporal behaviours. For example it is common for a user to buy a laptop, and then subsequently buy a laptop bag. But it is unusual for a user to buy a laptop bag first and then buy a laptop later. The challenge of this sequential approach is that it exacerbates underlying data scarcity issues. The number of available training samples is the limiting factor in most recommender systems, and the number of available interaction sequences is orders of magnitude less than the number of raw interactions.

On the other hand, integrating multimodal recommender techniques into our system, helps tackle the interaction scarcity problem from a different perspective. The use of additional modalities provides the opportunity for a much richer understanding of the items being recommended. It also becomes possible to infer relationships directly between similar and dissimilar items, beyond those present in the historical interaction data. 
By using Graph Neural Networks (GNNs) to model these relationships, we are also exploiting collaborative filtering techniques. This concretely reduces interaction scarcity, by sharing training examples across similar users/items.

In summary, by unifying advances from both multimodal and sequential recommender systems the proposed MuSTRec approach merges advantages from the three primary paradigms of recommendation systems: collaborative, content-based, and context-based filtering. These allow us to respectively: (1) leverage the user-item interaction history to identify patterns/similarities and share training constraints among related users or items, (2) exploit item attributes, such as textual descriptions and images features to inject additional relationships between items, beyond those present in the training interactions, (3) capture the temporal dynamics in user behaviours and provide recommendations relevant to the user's current situation. This holistic integration leads to better recommendation performance by maximizing the extracted training information.

In practice, MuSTRec includes both temporal and a-temporal recommendation heads, based on shared multimodal feature embedding graphs for items and users. These are trained in an end-to-end fashion.
We create a user-item bipartite graph, with noisy edges removed by degree-sensitive edge pruning \cite{zhou2023tale}. We also create item-item graphs for each data modality, which are frozen during training. The item embeddings are concatenated into sequences of embeddings based on the interaction sequences of users. 
A transformer-like network head operates on these embedding sequences.
To overcome oversmoothing issues, this transformer head integrates Fourier transforms and a frequency rescaler, applying a high pass filter to capture short-term shifts in user behaviour \cite{shin2024attentive}.

Extensive experiments are conducted on three real-world datasets and we compare against state-of-the-art multimodal and sequential recommendation methods. Our model significantly outperforms all baseline models across all metrics (up to 33.5\%), highlighting the value of this unified framework for future  recommendation systems.

\section{Related Work}
\label{sec:litrev}
\subsection{Multimodal Recommender Systems}
To incorporate multimodal information into the collaborative filtering scheme, Multimodal Recommender Systems (MMS) generally incorporate deep or graph learning methods. VBPR \cite{he2016vbpr} was one of the earliest works in this space, extending the classical BPR method \cite{rendle2012bpr}, by proposing a BPR based loss and incorporating image features into recommendation by concatenating image embeddings with item-ID embeddings. Attention mechanisms have also been used to discern user preferences for multimodal features \cite{zhou2023comprehensive}. 

Motivated by capturing high order semantics in user and item representation, recommender systems have seen a rise in GNN integration. For instance, the Multi-Modal Graph-Based Recommendation System \cite{gamal2023} integrates heterogeneous modalities of information using GNNs to achieve a better recommendation performance. MMGCN \cite{wei2019mmgcn} builds user-item bipartite graphs based on every modality. Representations from every modality graph are aggregated to get the final enhanced user and item representations. GRCN \cite{yu2021graph} builds on MMGCN by removing false-positive edges from the graphs before aggregation and message propagation is carried out. 

DualGNN \cite{wang2021dualgnn} introduces a user-user co-occurence graph constructed from modality-specific graph representations using a fine-grained spectral clustering method. This is done to capture the changing preferences of users for different modality features. This implicitly extracts item-item relations. On the other hand, the authors of LATTICE \cite{zhang2021mining} explicitly construct item-item relation graphs on each modality before fusing them together to obtain a joint multimodal item-item. This is dynamically updated during training by joint optimisation of the modality embeddings and the item-item adjacency matrix. The authors of FREEDOM \cite{zhou2023tale} observed that freezing the LATTICE item-item graph improves performance slightly. To this end, they propose a simpler framework that uses multimodal graphs, precomputed from raw multimodal data, that remain fixed during training. This enables information propagation in the GCN depending only on the graph structure. It also included additional BPR losses to align the user embeddings with modality features, reinforcing modality fusion. Additionally, degree-sensitive edge pruning is implemented to denoise the user-item graph getting rid of redundant edges. 

BM3 \cite{zhou2023bootstrap} introduces a self-supervised learning method, lowering computation cost and minimising impact of unreliable supervision signals. A straightforward dropout technique is employed to create contrastive views and three distinct contrastive loss functions optimise the representations. SLMRec \cite{tao2022self} incorporates self-supervised learning into GNNs to better capture underlying relationships. This is used alongside the supervised task to discover hidden data signals. Self-supervised data augmentation techniques are carried out, followed by optimisation via a contrastive loss. This combination of differently supervised heads for a unified model somewhat mirrors the MuSTRec approach we propose. However, we consider entirely different recommendation paradigms rather than simply different supervision sources.

\subsection{Sequential Recommender Systems}
Sequential recommender systems rely on user's historical interaction pattern to suggest the next item. Various research efforts have explored distinct methodologies to achieve this. 
Markov Chains were employed in earlier models to navigate item transitions \cite{rendle2010factorizing}. Other methods, like Caser \cite{tang2018personalized}, applied convolutional neural networks (CNNs) to capture local item relationships after treating the sequence of item embeddings as an image.

The rise in deep learning methods substantially affected sequential recommendation, leading to the utilisation of RNNs \cite{vaswani2017attention}. The use of Gated Recurrent Units (GRUs) was subsequently proposed in GRU4Rec \cite{hidasi2016}.
These recurrent approaches were limited by their sequential processing nature and vanishing gradient issues.
The success of transformer-based models in other fields 
inspired further exploration of self-attention for recommender systems \cite{li2024recent} to encode complex user behaviour patterns and capture long-range dependencies.
SASRec \cite{kang2018self} uses bidirectional attention, and illustrates the power of transformers where it is able to simultaneously learn past and future user actions.

Unfortunately, transformer-like models suffer from two key limitations. First, processing sequences with self-attention historically requires greater data volumes to compensate for the lack of inductive biases compared to structured models like convolution. Approaches like BERT4Rec \cite{sun2019} are restricted by this, leading to weaker generalisation.
To mitigate this, contrastive learning was utilised in DuoRec \cite{qiu2022contrastive}, combining unsupervised augmentation and supervised semantic positives. 

The second and more important issue with transformers, is the so-called ``oversmoothing problem''. Previous works \cite{du2023frequency} have observed that self-attention acts like a low-pass filter; transformer-like architectures generally smooth out fine detail in user behaviour patterns.
To solve this, contrastive learning was also used in FEARec \cite{du2023frequency}, which separates low and high frequency information and utilises frequency normalisation. FMLPRec \cite{fmlprec} employs a global filter to reduce noise in the frequency domain. While this global filter emphasises low frequencies, it has a tendency to underrepresent higher ones without large amounts of data. AC-TSR \cite{zhou2023attention} adjusts unreliable attention weights in Transformers and AdaMCT \cite{jiang2023adamct} utilises a local convolution filter to integrate locality-induced bias. 
BSARec \cite{shin2024attentive} uses a Fourier transform to provide the model direct insight from the frequency information. It also introduces a frequency scaler applying a high pass filter to limit oversmoothing. In MuSTRec we include a similar Fourier based representation specifically within the sequential detection head.

\section{Preliminaries}
\subsection{Item-Item Graphs}
Firstly, following \citet{zhou2023tale}, every modality present in the dataset is processed through a pre-trained modality-specific feature encoder. These produce the embedding vectors \( \mathbf{x}_i^m \) for modality $m$ of each item $i$.
Item-item graphs are created for each modality. To this end we construct the modality's similarity matrix \( S^m \), where the entry \( S_{ij}^m \) contains the Cosine Similarity between the embeddings of item pairs, \( \mathbf{x}_i^m \), \( \mathbf{x}_j^m \),
\begin{equation}
S_{ij}^m = \frac{\mathbf{x}_i^m \cdot \mathbf{x}_j^m}{\|\mathbf{x}_i^m\| \|\mathbf{x}_j^m\|}.
\end{equation}

We subsequently sparsify the graph, with only the top-k most similar items being retained for each item using kNN sparsification. We choose k to be 10, and create the sparsified matrix \( \hat{S}_{ij}^m \) as
\begin{equation}
\hat{S}_{ij}^m = 
\begin{cases} 
1, & \text{if } S_{ij}^m \in \text{top-}k(S_i^m) \\
0, & \text{otherwise}.
\end{cases}
\end{equation}
Next, \( \hat{S}_{ij}^m \) is normalised as \( \tilde{S}^m = (D^m)^{-\frac{1}{2}} \hat{S}^m (D^m)^{-\frac{1}{2}} \), where \( D^m \) is the degree matrix with \(D_{ii}^m = \sum_j \hat{S}_{ij}^m\). 

Finally, the combined multimodal item-item graph, $S$, is obtained through aggregation of modality-specific graphs: 

\begin{equation}
S = \sum_{m \in M} \alpha_m \tilde{S}^m,
\end{equation}

where \( \alpha_m \) is the weighting for modality \( m \), which is a hyperparameter of the model, and \( M \) is the set of all modalities. We consider only textual and visual information in our work similar to \cite{zhou2023tale, zhang2021mining}.

\subsection{User-Item Graphs}
Unlike the item-item graph, the user-item graph does not start from a feature similarity matrix. Rather we start from the binary adjacency matrix, $R$ where \( R_{ui} = 1 \) if user $u$ has interacted with item $i$. We then convert this to a bipartite graph $\hat{A}$ as 
\begin{equation}
\hat{A} = 
\begin{bmatrix}
0 & R \\
R^\top & 0
\end{bmatrix}
\end{equation}

We denoise the user-item graph by pruning edges connecting to popular nodes to limit oversmoothing following a degree-sensitive probability calculation \cite{zhou2023tale}. 
The edge retention probability is calculated by:
\begin{equation}
p_{ij} = \frac{1}{\sqrt{\omega_i} \sqrt{\omega_j}},
\end{equation}

where \( \omega_i \), \( \omega_j \) are the degrees of nodes \( i \) and \( j \). High-degree edges thus have a low sampling probability from the graph and are more likely to be pruned. Pruning followed by normalisation is carried out on every training epoch to obtain the denoised normalised adjacency matrix, \( A \):

\begin{equation}
A = D^{-\frac{1}{2}} \hat{A} D^{-\frac{1}{2}}
\end{equation}

\section{Methodology}\label{sec:method}
\begin{figure}[t]
\centering
\includegraphics[width=0.8\columnwidth]{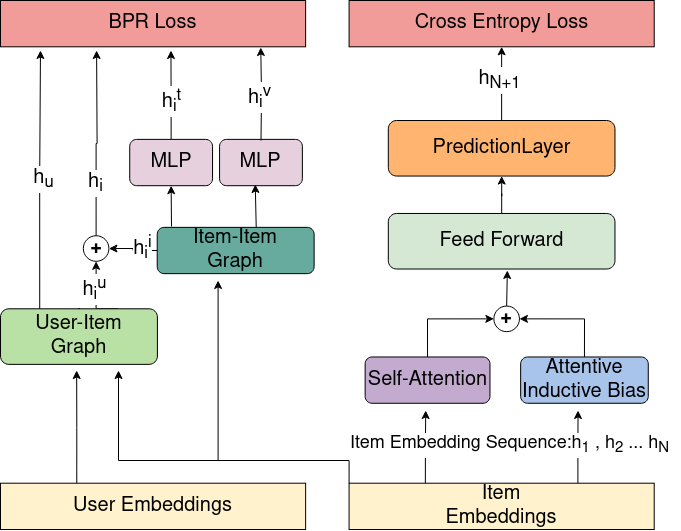}
\caption {The MuSTRec Architecture}
\label{fig1}
\end{figure}

As shown in Figure \ref{fig1}, MuST-Rec comprises of two major components, a GNN backbone and transformer head. Within the backbone, a series of Graph Convolution layers serve to align the user and item modality information, while also performing collaborative filtering. These embeddings are then reshaped into sequences and fed to a transformer based sequential prediction head, with frequency based filtering.

\subsection{Graph Embedding}
We implement graph convolutions using LightGCN \cite{he2020lightgcn}, and recursively define a convolution operating on a generic graph $X$ at layer $l$ as
\begin{equation}
G^{(l)}(X_i) = \sum_{j \in N(i)} X_{ij} G^{(l-1)}(X_j), 
\end{equation}
where at the base case $G^{(0)}(X_i) = X_i$.

We apply this definition to embed the user-item and item graphs through a series of \(L_{ui}\) and \(L_{ii}\) graph convolutions respectively:
\begin{equation}
\hat{h}_v = G^{(L_{ui})}(A_v) 
\end{equation}
\begin{equation}
\tilde{h}_i = G^{(L_{ii})}(S_i). 
\end{equation}

The final user embeddings $h_u$ are then simply extracted from the embedded user-item graph as \(h_u = \hat{h}_u\). Meanwhile, the final item embeddings $h_i$ are defined as the combination of that item's embedding within the item-item graph and the user-item graph, \(h_i = \tilde{h}_i + \hat{h}_i\).

\subsection{Sequential Prediction Head}
The time-agnostic item embeddings emerging from the graph-convolution network are subsequently reshaped into temporal interaction sequences and passed into a Transformer-like architecture that captures long and short term user preferences via frequency analysis, inspired by \cite{shin2024attentive}. Specifically, the input sequence, \(S_u\),  is composed of the embeddings of items interacted with by user $u$, arranged in temporal order
\begin{equation}
S_u = [h_{s_1}, h_{s_2}, \ldots, h_{s_{N-1}}].
\end{equation}
In Section \ref{sec:user_embed} of the experiments, we further explore the possibility of pre-pending this input sequence with the user-embedding (i.e. $S_u = [h_u, h_{s_1}, h_{s_2}, \ldots, h_{s_{N-1}}]$).

The user and item embeddings integrated into these sequences are significantly enriched compared to those traditionally used in sequential recommender systems. Not only do they include complex multimodal information, but the user-item and item-item graphs ensure that significant collaborative filtering has occurred, enabling the sharing of training information across similar items and users.

At this point, we further enrich the input sequence with positional embeddings to enable the transformer to learn the temporal dynamics which are omitted from the graph embedding process
\begin{equation}
E_u = \text{LayerNorm}(S_u + P).
\end{equation}

At block \(l\) of our transformer head, we define the input as \(X^l\), where in the base case \(X^0 = E_u\). The self-attention matrix for the block is then computed as follows:
\begin{equation}
\Lambda^{(l)} = \text{Softmax} \left( \frac{Q^{(l)} (K^{(l)})^\top}{d} \right),
\end{equation}
with the query and key embeddings defined as \(Q^{(l)} = X^{(l)} W_Q^{(l)}, \quad K^{(l)} = X^{(l)} W_K^{(l)}\) using \( W_Q^{(l)} \) and \( W_K^{(l)} \) learnable weight matrices respectively.

Following \citet{shin2024attentive}, we additionally define a frequency-based attention matrix, using Fourier transforms to incorporate inductive bias associated with low (LFC) and high frequency (HFC) user-item interaction patterns:
\begin{equation}
\Lambda_{IB}^{(l)} (X^{(l)}) = \text{LFC}[X^{(l)}] + \beta \text{HFC}[X^{(l)}],
\end{equation}
where \(\beta\) is a trainable parameter adjusting the emphasis of the high frequency components.

We subsequently apply both types of self-attention to the signal, and blend the results according to the hyperparameter \(\alpha\):
\begin{equation}
S^{(l)} = \alpha \Lambda_{IB}^{(l)} X^{(l)} + (1 - \alpha) \Lambda^{(l)} X^{(l)}
\end{equation}

We subsequently perform a two-layer feed-forward projection using learnable weight matrices \(W_1^{(l)}\) and \(W_2^{(l)}\) and a GELU activation:
\begin{equation}
\tilde{X}^{(l)} = \text{GELU}(S^{(l)} W_1^{(l)} + b_1^{(l)}) W_2^{(l)} + b_2^{(l)}
\end{equation}

Finally, dropout, residual connections and normalisation are incorporated to provide the full recursive definition of transformer block $l$:
\begin{equation}
X^{(l+1)} = \text{LayerNorm}(X^{(l)} + \text{Dropout}(\tilde{X}^{(l)}))
\end{equation}

The output from the final Transformer head layer $L$ is treated as the embedding of the next item this user is predicted to interact with. To make this embedding concrete, the cosine similarity of the resulting output is computed against all the item embeddings
\begin{equation}
\hat{y}_i = h_i^\top X_{N}^{(L)}.
\end{equation}
These preference scores can be sorted and truncated to produce the final list of candidate items.

\subsection{Training and Optimisation}
Two losses are proposed to optimize the model. Firstly the 
Bayesian Personalized Ranking (BPR) Loss is utilised for optimising the graph-based embeddings, encouraging the model to rank interacted items $i$ higher than non-interacted ones $j$. This is applied both to the combined user-item graph and to the independent per-modality item embeddings
\begin{align}
L_{\text{BPR}} = \sum_{(u, i, j) \in D} \bigg( & -\log \sigma(h_u^\top h_i - h_u^\top h_j) \nonumber \\
& + \lambda \sum_{m \in M} -\log \sigma(h_u^\top h_i^m - h_u^\top h_j^m) \bigg).
\end{align}
This is a regularisation, which helps improve the quality of the embeddings without directly referring to the next interaction in the sequence.

We additionally constrain the model by treating the final similarity score outputs $\hat{y}$ as the logits of the transformer head, and computing a Cross-Entropy loss against the ground truth $g$ next item in the sequence
\begin{equation}
L_{\text{CE}} = -\log \frac{\exp(\hat{y}_g)}{\sum_{i \in V} \exp(\hat{y}_i)}
\end{equation}

A hyperparameter \(\omega\) balances the two losses to form the final objective function:
\begin{equation}
L_{\text{total}} = L_{\text{BPR}} + \omega L_{\text{CE}}
\label{eq:loss}
\end{equation}

We also introduce a pretrained variant of our model, MuSTRec-S. In this variant, the user-item and item-item graphs are first trained to generate embeddings with \(L_{BPR}\). These graphs are then frozen before passing item embeddings to the sequential head.

\section{Experiments}

We conduct several experiments to evaluate our proposed model and determine how it compares to the previous state-of-the-art sequential and multimodal recommender systems. 
Importantly, the prevalent practice in the fields of sequential and multimodal recommendation are similar but not identical (e.g. the separation of data between the train, validation and test sets, or the exact definition of particular metrics). Therefore, although we use standard datasets, we find that we must develop a new more nuanced evaluation protocol to enable these comparisons. 

Beyond comparisons to prior works, we also carry out an ablation study to assess the contribution of the different components of our system. Finally, we determine the sensitivity of MuSTRec with respect to \(\omega\) to evaluate how the performance on the different metrics varies by changing the contribution of the multimodal and sequential components.

\subsection{Experimental Setup}

Traditional multimodal recommendation systems are atemporal and use random train/test splits. This doesn't mesh well with sequential recommendation, because the prediction target may be in the past.
We define a new evaluation protocol where the data is split such that all historical interactions except the last two are used for training, the second to last is for validation, and the final one for testing. This split methodology is employed for sequential and multimodal recommender systems to ensure an equitable comparison. Because this is the first work to unify multimodal and sequential recommendation evaluation, we reran all the baseline experiments using these splits. All possible subsequences are created from every user's training interactions and one is randomly selected during each training iteration. The entire available sequence is used during validation and testing to evaluate the model's performance on most recent interactions.

Since each user has exactly one item in both the validation and test sets, Hit Rate@k (HR@k) indicates whether that single ground-truth item appears in the top-k recommendations, making it a suitable metric for both sequential and multimodal baselines with these data splits. We also report Normalised Discounted Cumulative Gain (NDCG@k) to assess the ranking quality of the recommendations.

\subsection{Datasets}

\begin{table}[htbp]
\centering
\caption{Datasets Statistics}
\vskip 0.1in
\label{tab:amazon_5core_stats}
\begin{tabular}{lrrrr}
\toprule
Dataset & Users & Items & Reviews & Sparsity (\%) \\
\midrule
\textit{Sports}      & 35{,}598   & 18{,}357   & 296{,}337   & 99.95 \\
\textit{Clothing}    & 39{,}387   & 23{,}033   & 278{,}677   & 99.97 \\
\textit{Elec}        & 192{,}403  & 63{,}001   & 1{,}689{,}188 & 99.99 \\
\textit{Baby}        & 19{,}445   & 7{,}333    & 160{,}792   & 99.89 \\
\bottomrule
\end{tabular}
\end{table}

The popular and publicly available collection of Amazon datasets, contain product reviews extracted from Amazon.com \cite{mcauley2014amazon}. They are widely used to evaluate models across sequential and multimodal recommendation tasks. The three datasets of ``Sports and Outdoors'', ``Clothing, Shoes and Jewelry'', ``Electronics'' and ``Baby'' are explored in detail here (denoted as \(Sports\), \(Clothing\) and  \(Elec\), and \(Baby\), respectively, for simplicity). Table \ref{tab:amazon_5core_stats} shows the statistics of these four datasets.

The text description and URL of images that could be utilised as textual and visual modalities are contained in the metadata of the Amazon datasets.
We use the published, 384-dimensional text features, extracted using pre-trained sentence-transformers \cite{reimers2019sentence} and 4096-dimensional visual features by following \cite{zhang2021mining}. We build the user-item graph based on the review ratings after obtaining the textual and visual embeddings and treating the user ratings as a positive interaction.

\subsection{Baselines}

We compare our model with several state-of-the-art methods for multimodal and sequential recommendation. These fall under three categories:
\begin{itemize}
    \item Two generic collaborative filtering techniques, BPR \cite{rendle2012bpr} and LightGCN \cite{he2020lightgcn}, relying on user-item interactions only
    \item Six multimodal recommender systems, FREEDOM \cite{zhou2023tale}, LGMRec \cite{guo2024lgmrec}, VBPR \cite{he2016vbpr}, MMGCN \cite{wei2019mmgcn}, MGCN \cite{yu2023multi} and GRCN \cite{yu2021graph}, which have shown superior performance when compared to other multimodal recommender systems \cite{zhou2023comprehensive}. These models rely on interaction data as well as multimodal item data for accurate recommendations
    \item Five state-of-the-art sequential models, BSARec \cite{shin2024attentive}, BERT4Rec \cite{sun2019}, FEARec \cite{du2023frequency}, SASRec \cite{kang2018self} and FMLPRec \cite{fmlprec}. These models take sequences of user interactions to predict the next item, capturing temporal user dynamics.
\end{itemize}

To re-iterate, all of these baselines were re-evaluated under the new unified evaluation protocol specified above.

\subsection{Implementation Details}
\begin{table}[htbp]
\centering
\caption{Performance Metrics Across Datasets. Type indicates if a technique is \textbf{M}ultimodal or \textbf{S}equential. All metric values are displayed as percentage (multiplied by 100).}
\label{tab:results}
\begin{tabular}{p{1.9cm}p{0.4cm}p{0.9cm}p{0.9cm}p{0.9cm}p{0.9cm}}
\toprule
Model & Type & HR@10 & HR@20 & N@10 & N@20 \\
\midrule
\(Baby\) &  & & & & \\
BSARec & S & 5.08 & 7.60 & 2.82 & 3.46 \\
BERT4Rec & S & 4.06 & 6.48 & 2.06 & 2.67 \\
FEARec & S & 4.82 & 7.36 & 2.60 & 3.24 \\
SASRec & S & 3.10 & 5.11 & 1.63 & 2.13 \\
FMLPRec & S & 3.13 & 5.27 & 1.57 & 2.11 \\
FREEDOM & M & 4.09 & 6.80 & 2.09 & 2.77 \\
LGMRec & M & 4.23 & 6.73 & 2.21 & 2.83 \\
VBPR & M & 2.92 & 4.79 & 1.50 & 1.96 \\
MMGCN & M & 2.91 & 4.64 & 1.40 & 1.84 \\
MGCN & M & 4.14 & 6.54 & 2.20 & 2.81 \\
GRCN & M & 3.81 & 5.93 & 1.99 & 2.52 \\
LightGCN & - & 3.43 & 5.58 & 1.77 & 2.32 \\
BPR & - & 2.59 & 4.38 & 1.33 & 1.78 \\
\textbf{MuSTRec-S} & \textbf{MS} & \textbf{6.15} & \textbf{9.39} & \textbf{3.46} & \textbf{4.27}\\
\textbf{MuSTRec} & \textbf{MS} & \textbf{5.89} & \textbf{8.51} & \textbf{3.32} & \textbf{3.98} \\
\midrule
\(Clothing\) &  & & & & \\
  BSARec & S &      3.43 &      4.78 &        1.96 &        2.29 \\
BERT4Rec & S &      1.91 &      3.01 &        0.98 &        1.26 \\
  FEARec & S &      3.11 &      4.33 &        1.80 &        2.11 \\
  SASRec & S &      1.67 &      2.58 &        0.86 &        1.09 \\
 FMLPRec & S &      1.91 &      2.93 &        1.02 &        1.27 \\
 FREEDOM & M &      4.38 &      6.78 &        2.30 &        2.90 \\
  LGMRec & M &      3.68 &      5.73 &        1.92 &        2.44 \\
    VBPR & M &      2.57 &      4.04 &        1.37 &        1.74 \\
   MMGCN & M &      1.67 &      2.78 &        0.82 &        1.10 \\
    MGCN & M &      4.59 &      6.78 &        2.40 &        2.95 \\
    GRCN & M &      3.06 &      4.76 &        1.62 &        2.05 \\
LightGCN & - &      2.90 &      4.45 &        1.53 &        1.91 \\
     BPR & - &      1.87 &      2.80 &        1.00 &        1.23 \\
\textbf{MuSTRec-S} & \textbf{MS} & \textbf{6.19} & \textbf{8.33} & \textbf{3.36} & \textbf{4.02}\\
\textbf{MuSTRec} & \textbf{MS} & \textbf{5.19} & \textbf{7.57} & \textbf{2.79} & \textbf{3.39} \\
\midrule
\(Sports\) &  & & & & \\
BSARec & S & 5.56 & 7.93 & 3.28 & 3.88 \\
BERT4Rec & S & 3.39 & 5.30 & 1.78 & 2.26 \\
FEARec & S & 5.49 & 7.84 & 3.21 & 3.80 \\
SASRec & S & 2.80 & 4.29 & 1.46 & 1.83 \\
FMLPRec & S & 3.31 & 5.02 & 1.78 & 2.21 \\
FREEDOM & M & 4.81 & 7.55 & 2.51 & 3.20 \\
LGMRec & M & 4.66 & 7.40 & 2.45 & 3.15 \\
VBPR & M & 4.06 & 6.38 & 2.04 & 2.63 \\
MMGCN & M & 2.86 & 4.62 & 1.41 & 1.86 \\
MGCN & M & 4.82 & 7.57 & 2.55 & 3.25 \\
GRCN & M & 3.97 & 6.29 & 2.07 & 2.65 \\
LightGCN & - & 4.02 & 6.36 & 2.12 & 2.71 \\
BPR & - & 3.17 & 4.98 & 1.63 & 2.08 \\ 
\textbf{MuSTRec-S} & \textbf{MS} & \textbf{6.88} & \textbf{9.96} & \textbf{3.82} & \textbf{4.59}\\
\textbf{MuSTRec} & \textbf{MS} & \textbf{6.19} & \textbf{8.88} & \textbf{3.39} & \textbf{4.06} \\
\bottomrule
\end{tabular}
\end{table}

\begin{table}[h]
    \centering
    \caption{\(Elec\) performance comparison results. Type indicates if a technique is \textbf{M}ultimodal or \textbf{S}equential. All metric values are multiplied by 100}
    \vskip 0.1in
    \begin{tabular}{p{1.9cm}p{0.4cm}p{0.9cm}p{0.9cm}p{0.9cm}p{0.9cm}}
\toprule
Model & Type & HR@10 & HR@20 & N@10 & N@20\\
\midrule
BSARec & S & 5.46 & 7.64 & 3.17 & 3.72 \\
BERT4Rec & S & 4.06 & 6.03 & 2.20 & 2.69 \\
FEARec & S & 5.35 & 7.52 & 3.11 & 3.66 \\
SASRec & S & 2.81 & 4.27 & 1.46 & 1.83 \\
FMLPRec & S & 2.99 & 4.52 & 1.55 & 1.94 \\
FREEDOM & M & 3.25 & 4.93 & 1.76 & 2.18 \\
LGMRec & M & 3.21 & 4.82 & 1.72 & 2.12 \\
VBPR & M & 2.28 & 3.48 & 1.13 & 1.43 \\
MMGCN & M & 2.01 & 3.19 & 1.03 & 1.33 \\
MGCN & M & 3.32 & 4.98 & 1.80 & 2.21 \\
GRCN & M & 2.12 & 3.40 & 1.12 & 1.45 \\
LightGCN & - & 2.97 & 4.42 & 1.61 & 1.97 \\
BPR & - & 2.41 & 3.75 & 1.23 & 1.57 \\
\textbf{MuSTRec-S} & \textbf{MS} & \textbf{5.57} & \textbf{7.99} & \textbf{} & \textbf{3.78}\\
\textbf{MuSTRec} & MS & \textbf{5.50} & \textbf{7.72} & \textbf{3.19} & \textbf{3.75} \\
\bottomrule
    \end{tabular}
    \label{tab:elec_comp}
\end{table}



We train all models in PyTorch with Adam, Xavier initialisation and 64‑dim embeddings. Hyper‑parameters sweeps follow each paper’s recommendations; we fix the number of GCN layers in the item-item graph at $L_{ii}=1$ and the user-item bipartite graph at $L_{ui}=2$, \(\lambda=0.001\), visual feature ratio 0.1, early‑stopping 20 and 1000 total epochs, selecting on validation HR@20, in accordance with \cite{zhang2021mining}. For MuSTRec’s transformer encoder we employ two frequency-aware blocks (depth = 2) with 2-head attention (32-d per head), hidden size 64, max sequence length 50, 0.5 dropout on both attention and hidden/output layers, and no weight decay. Experiments run on a single NVIDIA 3090. MuSTRec and multimodal baselines were implemented within the unified recommendation framework, MMRec, to guarantee an equitable comparison \cite{zhou2023comprehensive}. Sequential baselines were incorporated into the unified framework \cite{shin2024attentive}.

\subsection{Performance Comparison}

Table \ref{tab:results} compares the recommendation performance of MuSTRec against all baseline models, in terms of HR and NDCG (Table \ref{tab:elec_comp} shows results on \(Elec\)). MuSTRec demonstrates improvements over BSARec, which is the best performing baseline across all datasets except \(Clothing\), using a similar transformer backbone to MuSTRec. On \(Baby\), MuSTRec achieves an increase of 15.94\% on HR@10, 11.97\% on HR@20, 17.73\% on NDCG@10 and 15.03\% on NDCG@20. Similarly, on \(Sports\), we observe an improvement of 11.33\%, 11.98\%, 3.35\% and 4.64\% on HR@10, HR@20, NDCG@10 and NDCG@20, respectively. The gain on \(Elec\) is more modest with a 0.73\%, 1.05\%, 0.63\% and 0.81\% improvement across HR@10, HR@20, NDCG@10 and NDCG@20, respectively. This may be because the Elec dataset is by far the largest (around 2 orders of magnitude larger than the others) and so the proposed solutions to data-sparsity are less impactful. MGCN is best performing baseline model on \(Clothing\) and MuSTRec achieves an improvement of 13.07\%, 11.65\%, 16.25\% and 14.92\% on HR@10, HR@20, N@10 and N@20, respectively. MuSTRec-S surpasses the best baseline by 22.68\% on \(Baby\), 26.89\% on \(Sports\), 33.50\% on \(Clothing\) and 1.97\% on \(Elec\), highlighting the benefit of pretraining the user-item and item-item graphs. It also outperforms MuSTRec by 10.69\% across all metrics and datasets.

On average, MuSTRec outperformed the best baselines by 9.4\% across all datasets and by 12.3\% when excluding the \(Elec\) dataset. The improvements can be attributed to the introduction of denoised user-item and frozen item-item embeddings into the sequences passed to the transformer-like encoder. 
Sequential models appear to perform better than multimodal models except on \(Clothing\). Considering only multimodal recommender systems, the best performing baseline is MGCN, but in this case MuSTRec outperforms it by an average of 40.00\% across all metrics and datasets.

\subsection{Ablation Study}

We decouple the user-item and item graphs from our transformer-like encoder and then we reintroduce one component at a time to evaluate their contribution to the recommendation accuracy. We define the following model variants as part of our ablation study: 
\begin{itemize}
    
\item\textbf{MuSTRec-B - } sequential-only baseline, \\
 \item\textbf{MuSTRec-M - } adds only the multimodal item-item graph, \\
 \item\textbf{MuSTRec-I - } adds only the user-item graph, \\
 \item\textbf{MuSTRec-V - } sets \( \alpha_m \) to only consider the contribution from visual data. \\
 \item\textbf{MuSTRec-T - } sets \( \alpha_m \)to only evaluate the textual information contribution.
 
\end{itemize}

\begin{figure}[htbp]
    \centering
    \includegraphics[width=0.9\columnwidth]{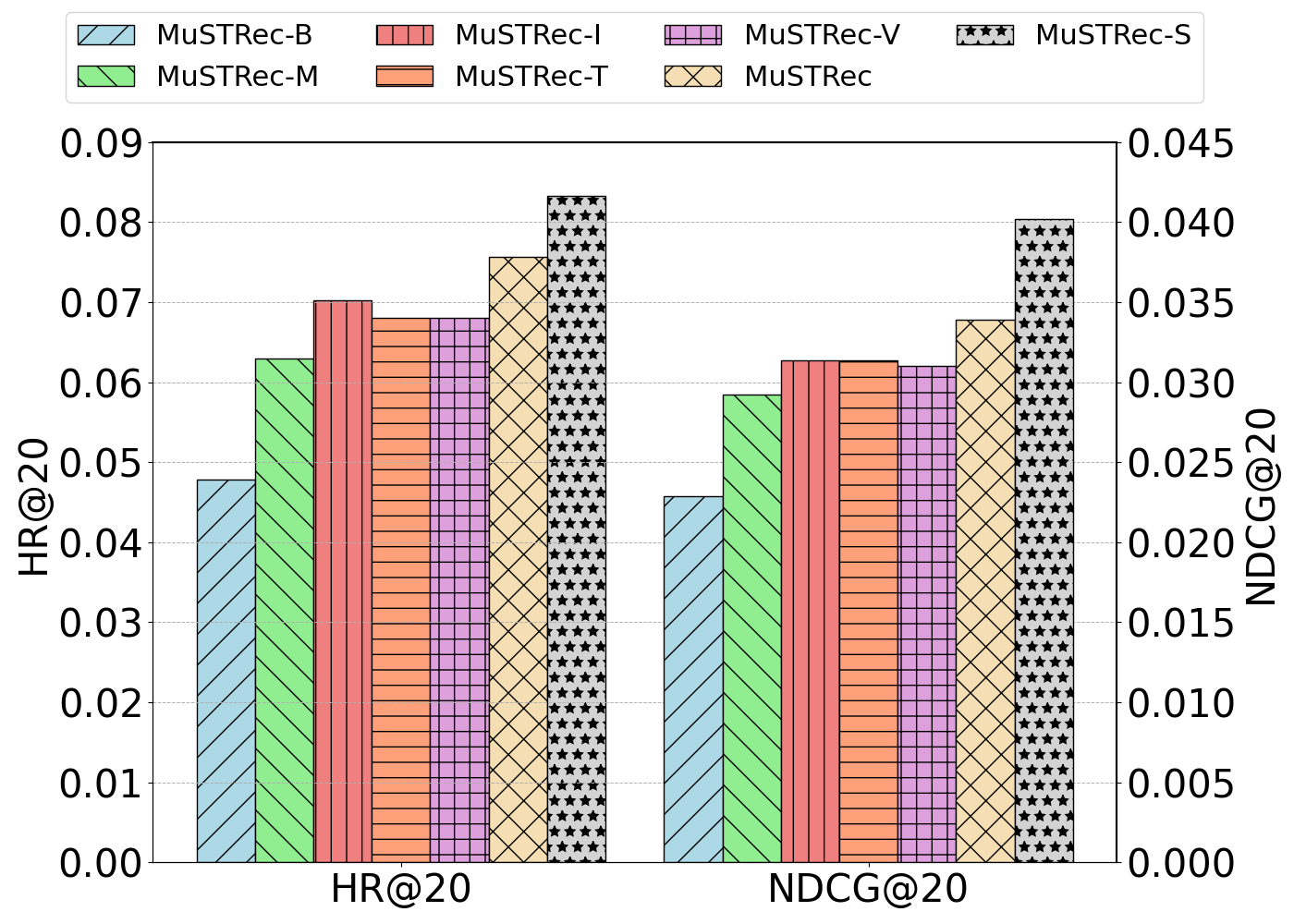}
    \caption{\(Clothing\) Ablation
    }
    \label{fig:clothing_ablation}
\end{figure}

\begin{figure*}[htbp]
\centering
\subfloat[\(Sports\) Sensitivity]{
    \includegraphics[width=0.32\linewidth]{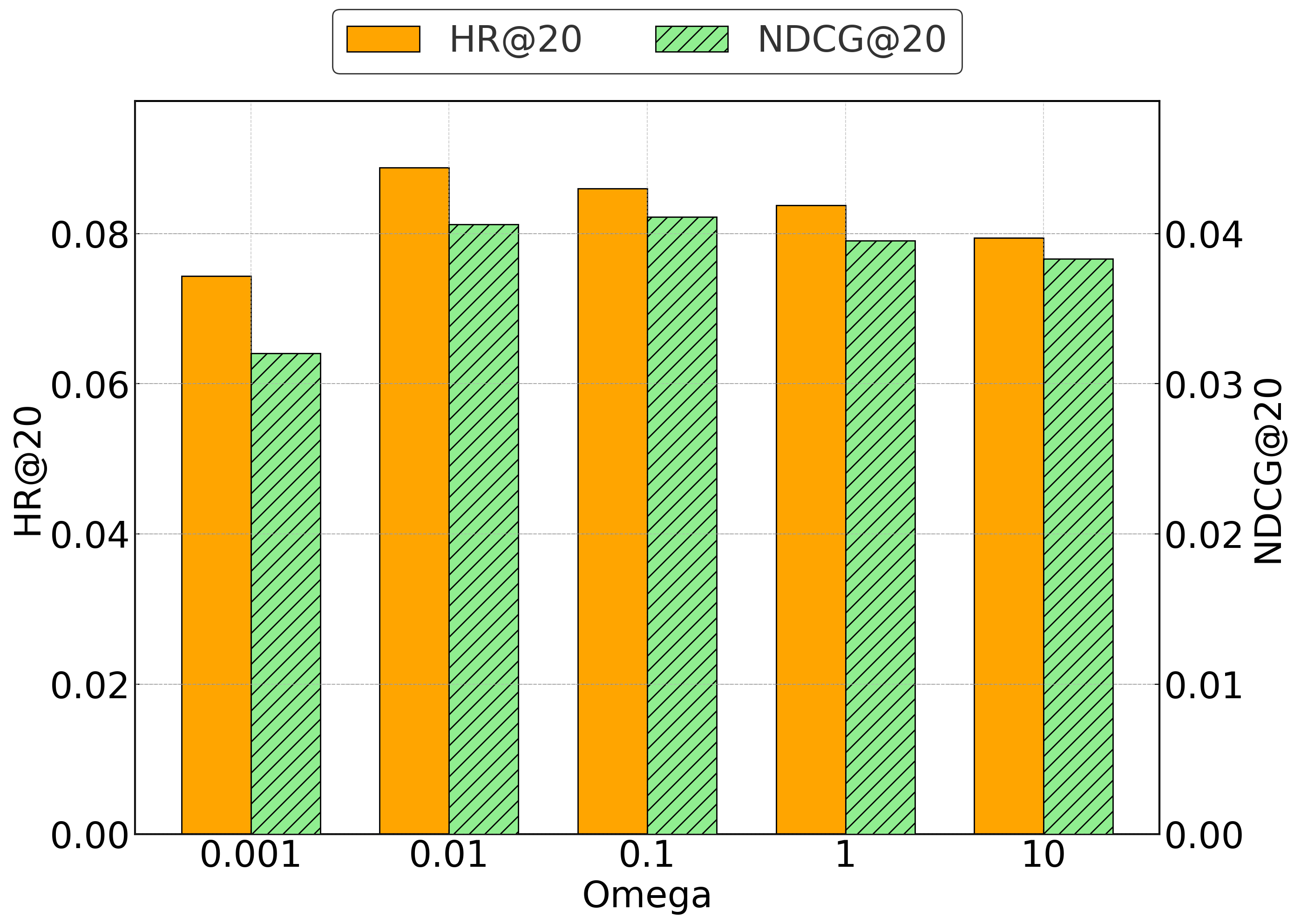}
    \label{fig:sports_sens}
}
\hfill
\subfloat[\(Clothing\) Sensitivity]{
    \includegraphics[width=0.32\linewidth]{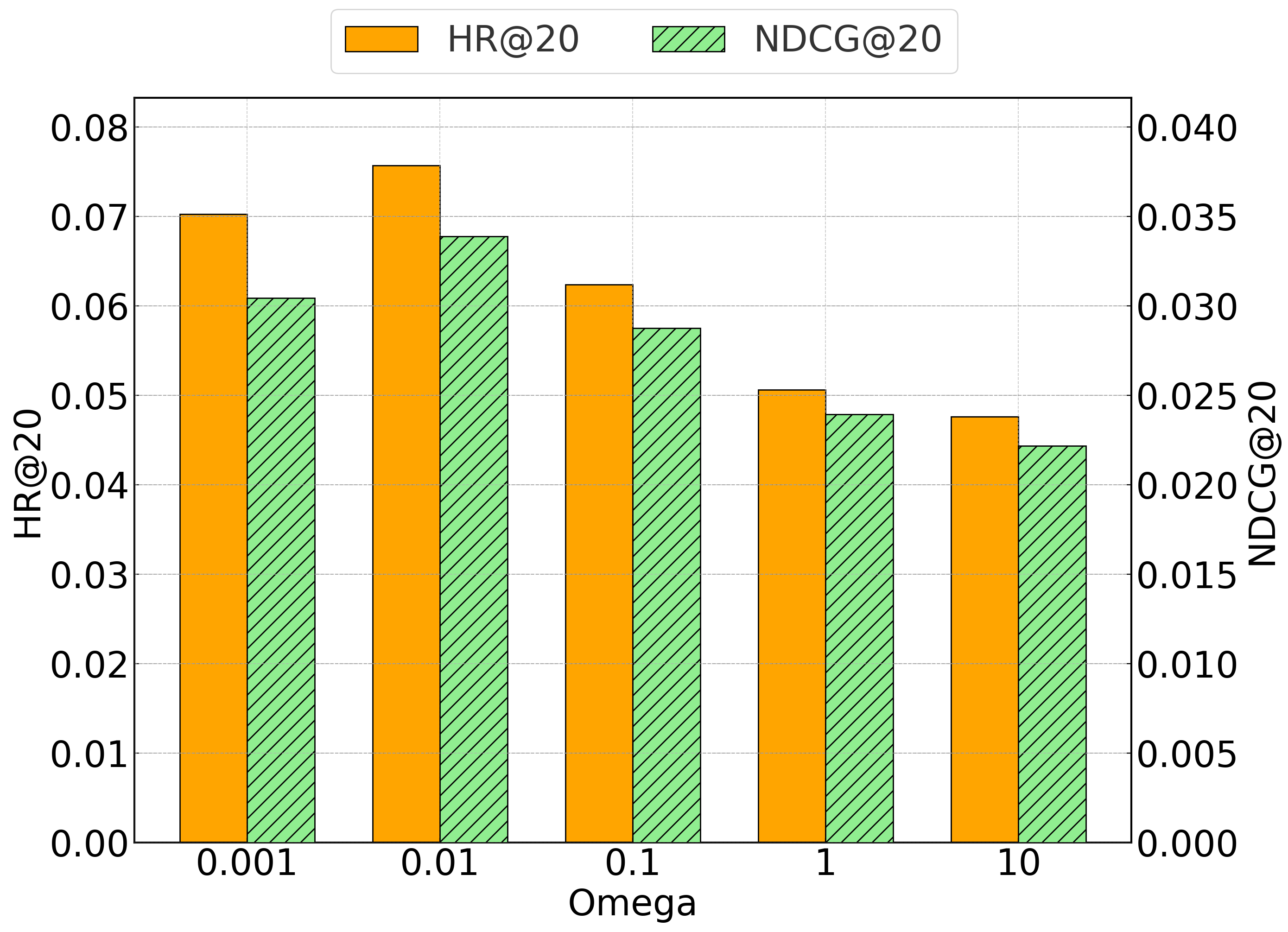}
    \label{fig:clothing_sensitivity}
}
\hfill
\subfloat[\(Elec\) Sensitivity]{
    \includegraphics[width=0.32\linewidth]{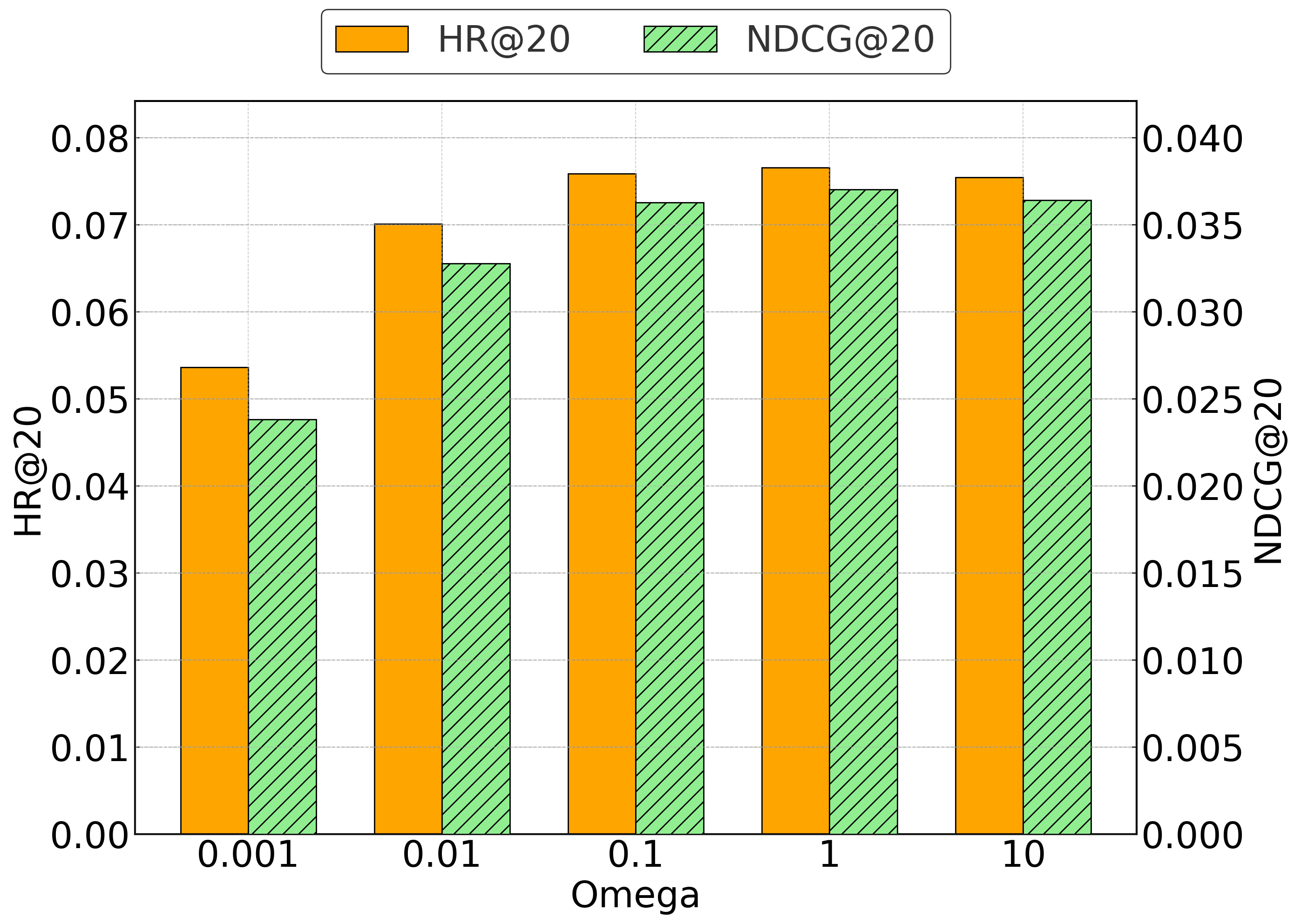}
    \label{fig:elec_sensitivity}
}

\vspace{-0.2cm}
\caption{Sensitivity analysis results on $\omega$ for the indicated datasets.}
\label{fig:combined_sensitivity_ablation}
\end{figure*}

Ablation results on \(Clothing\) in Figure \ref{fig:clothing_ablation} show that the individual components 
boost performance over the base model, with MuSTRec providing further gains through their combination and MuSTRec-S achieving the highest performance. We note that incorporating a user-item graph, as in MuSTRec-I, still improves performance over the sequential baseline, which could prove useful in applications where multimodal data may not be available.
For completeness, we include ablation results on the other datasets as Figures \ref{fig:baby_ablation}, \ref{fig:sports_ablation} and \ref{fig:elec_ablation} in Appendix A, where similar gains are observed from incorporating multimodal item-item graphs and pretraining.

\subsection{Sensitivity Study}

\begin{figure}[!htbp]
\centering
\includegraphics[width=0.9\columnwidth]{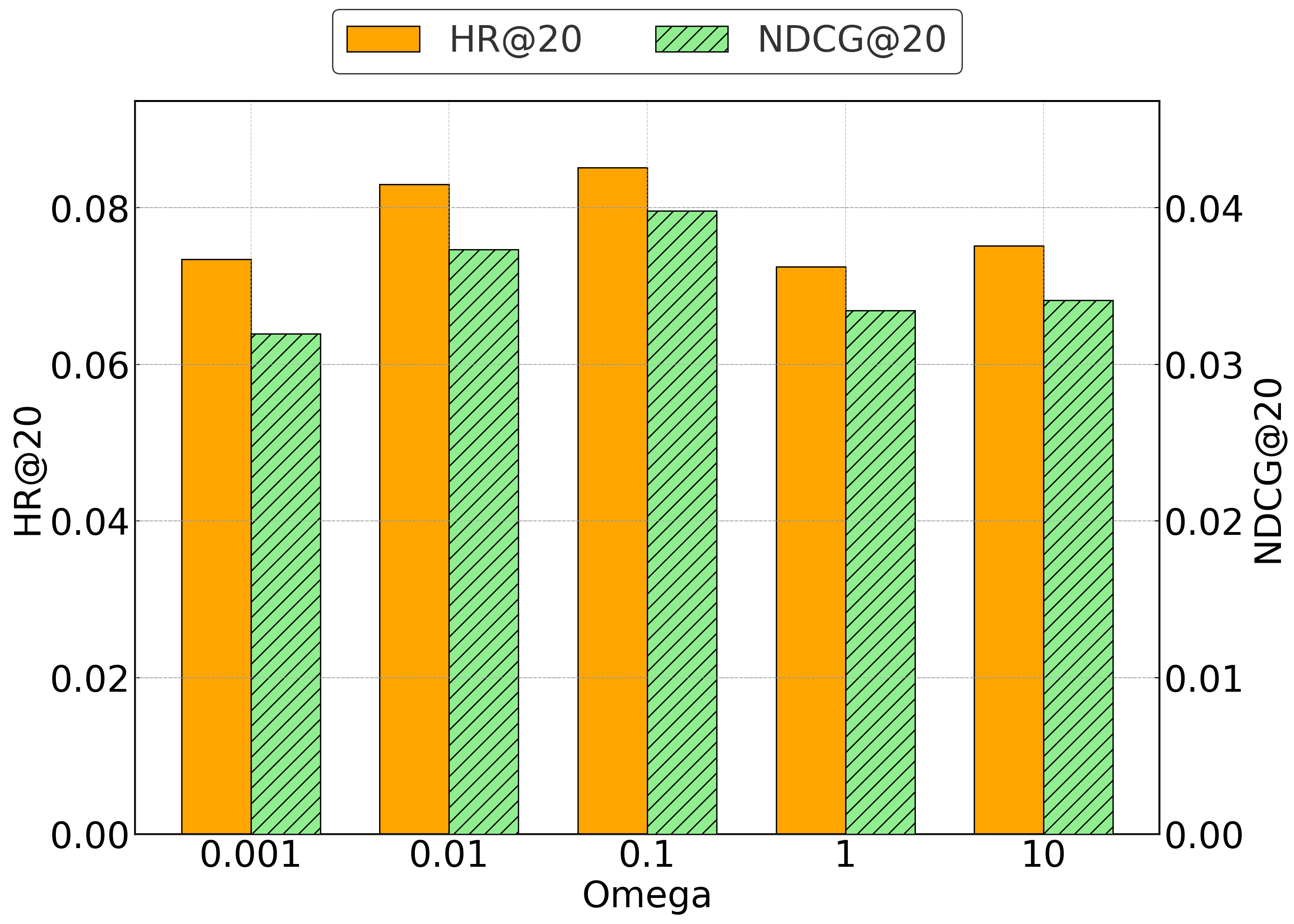}
\caption{\(Baby\) Sensitivity}
\label{fig:baby_sensitivity}
\end{figure}

In this subsection, we conduct sensitivity analysis with the hyperparameter \(\omega\), which controls the contribution of the sequential head component relative to the item-item and user-item graphs. We test $\omega$ in $\{0.001, 0.01, 0.1, 1, 10\}$. Figure \ref{fig:baby_sensitivity} (and \ref{fig:sports_sens}, \ref{fig:clothing_sensitivity}, \ref{fig:elec_sensitivity}) show N@20 and HR@20 performance metrics for different \(\omega\). We can observe that the best HR is achieved with \(\omega=0.1\) on \(Baby\), \(\omega=0.01\) on \(Sports\) and \(Clothing\) and \(\omega=1\) on \(Elec\). We can also note that sensitivity is generally low, with most hyperparameter settings achieving broadly similar results.

\subsection{Computation Cost}
MuSTRec sits roughly in the middle of the pack, with its training and inference times and GPU memory utilisation between the quickest and slowest models. 

\begin{table}[htbp]
\caption{Training/testing epoch times (seconds) and GPU memory usage for each model on \(Baby\).}
\centering
\begin{tabular}{lrrr}
\hline
\textbf{Model} & \textbf{Train time} & \textbf{Test time} & \textbf{GPU (MB)} \\
\hline
MuSTRec  & 30.5053 & 8.1470  & 1\,657.34 \\
BSARRec  & 9.3249  & 19.4332 & 1\,034.59 \\
BERT4Rec & 32.4977 & 21.1825 & 1\,423.26 \\
FEARec   & 142.9599 & 20.6301 & 5\,126.11 \\
FREEDOM  & 8.0681  & 2.7456 & 783.03 \\
LGMRec   & 15.7573 & 2.9911 & 427.26 \\
MMGCN    & 15.0671 & 2.7745 & 972.89 \\
LATTICE  & 7.7409  & 2.9352 & 2\,451.81 \\
\hline
\end{tabular}

\label{tab:model_perf}
\end{table}

\section{Impact of Concatenating User Embeddings}
\label{sec:user_embed}

\begin{table}[ht]
\caption{Overview of Validation (V) and Test (T) results on BSARec, MuSTRec, and MuSTRec-U (only @20).}
\centering
\setlength{\tabcolsep}{2pt}
\resizebox{\linewidth}{!}{%
\begin{tabular}{llr r r r r}
\toprule
\textbf{Dataset} & \textbf{Model} & \(\boldsymbol{\omega}\) 
& \textbf{V\_HR@20} & \textbf{T\_HR@20} & \textbf{V\_N@20} & \textbf{T\_N@20} \\
\midrule
\textit{Baby}   & BSARec    & -      & 0.0944 & 0.0760 & 0.0444 & 0.0346 \\
\textit{Baby}   & MuST-Rec  & 0.001  & 0.0959 & 0.0734 & 0.0414 & 0.0319 \\
\textit{Baby}   & MuST-Rec  & 0.010  & 0.1096 & 0.0830 & 0.0491 & 0.0373 \\
\textit{Baby}   & MuST-Rec  & 0.100  & 0.1082 & 0.0851 & 0.0501 & 0.0398 \\
\textit{Baby}   & MuST-Rec  & 1.000  & 0.0946 & 0.0725 & 0.0437 & 0.0334 \\
\textit{Baby}   & MuST-RecU & 0.001  & 0.3106 & 0.1904 & 0.1943 & 0.1191 \\
\textit{Baby}   & MuST-RecU & 0.010  & 0.2857 & 0.1741 & 0.1552 & 0.0959 \\
\textit{Baby}   & MuST-RecU & 0.100  & 0.1165 & 0.0781 & 0.0522 & 0.0341 \\
\textit{Baby}   & MuST-RecU & 1.000  & 0.0769 & 0.0560 & 0.0347 & 0.0243 \\
\midrule
\textit{Sports} & BSARec    & -      & 0.0988 & 0.0793 & 0.0480 & 0.0388 \\
\textit{Sports} & MuST-Rec  & 0.001  & 0.1050 & 0.0744 & 0.0466 & 0.0320 \\
\textit{Sports} & MuST-Rec  & 0.010  & 0.1158 & 0.0888 & 0.0534 & 0.0406 \\
\textit{Sports} & MuST-Rec  & 0.100  & 0.1065 & 0.0860 & 0.0514 & 0.0411 \\
\textit{Sports} & MuST-Rec  & 1.000  & 0.0998 & 0.0838 & 0.0481 & 0.0395 \\
\textit{Sports} & MuST-RecU & 0.001  & 0.2603 & 0.1467 & 0.1490 & 0.0789 \\
\textit{Sports} & MuST-RecU & 0.010  & 0.1958 & 0.1141 & 0.0954 & 0.0530 \\
\textit{Sports} & MuST-RecU & 0.100  & 0.1065 & 0.0704 & 0.0486 & 0.0302 \\
\textit{Sports} & MuST-RecU & 1.000  & 0.0785 & 0.0540 & 0.0363 & 0.0236 \\
\midrule
\textit{Elec}   & BSARec    & -      & 0.0845 & 0.0764 & 0.0406 & 0.0372 \\
\textit{Elec}   & MuST-Rec  & 0.001  & 0.0616 & 0.0536 & 0.0276 & 0.0238 \\
\textit{Elec}   & MuST-Rec  & 0.010  & 0.0792 & 0.0701 & 0.0369 & 0.0328 \\
\textit{Elec}   & MuST-Rec  & 0.100  & 0.0839 & 0.0759 & 0.0399 & 0.0363 \\
\textit{Elec}   & MuST-Rec  & 1.000  & 0.0829 & 0.0766 & 0.0400 & 0.0370 \\
\textit{Elec}   & MuST-RecU & 0.001  & 0.1270 & 0.0710 & 0.0652 & 0.0350 \\
\textit{Elec}   & MuST-RecU & 0.010  & 0.1117 & 0.0670 & 0.0570 & 0.0329 \\
\textit{Elec}   & MuST-RecU & 0.100  & 0.0749 & 0.0519 & 0.0346 & 0.0232 \\
\textit{Elec}   & MuST-RecU & 1.000  & 0.0633 & 0.0485 & 0.0293 & 0.0220 \\
\bottomrule
\end{tabular}}
\label{tab:MuSTRec-U_complete_20only}
\end{table}

A reasonable extension to the MuSTRec system would be to incorporate the learned user embeddings from the user-item graph into the sequential recommender head. In this section, we experimented with adding a special token to the start of every interaction sequence, which represented this user-specific embedding. This technique is referred to as MuSTRec-U.
MuSTRec-U is the model variant designed to integrate user embeddings into embedding sequences to explicitly incorporate user-specific information.

Surprisingly, we find that such an approach leads to an enormous increase in all metrics on the smaller datasets (Baby and Sports). In many cases, this leads to MuSTRec achieving accuracy improvements of 100\% or even 200\% over the best performing state-of-the-art baseline. An overview of results is provided in Table \ref{tab:MuSTRec-U_complete_20only}.
We found this result suspicious and conducted a battery of tests to investigate the phenomenon.

Table \ref{tab:baby_smaller_omega} shows that lower \(\omega\) values increase this effect, with the highest performance observed at \(\omega = 1e-05\). These very small \(\omega\) values weaken the training of the transformer head relative to the GNN embedding network constraints.  It is also noteworthy that there is a widened performance gap between the validation (one-step-ahead) and test (two-steps-ahead). Collectively, these might suggest that this regime relies heavily on the GNN to cluster users and their related items within the shared space, without considering the temporal aspects of the prediction problem.

\begin{table}[htbp]
\caption{Results for MuSTRec-U model on Baby dataset for smaller \(\omega\) values}
\centering
\small
\begin{tabular}{lccccc}
\hline
\textbf{Metric}  & \textbf{\(\omega=1e^{-04}\)} & \textbf{\(\omega=1e^{-05}\)} & \textbf{\(\omega=1e^{-06}\)} \\
\hline
Valid HR@10 & 0.2700 & 0.2712 & 0.2467 \\
Test HR@10 & 0.1628 & 0.1604 & 0.1506 \\
Valid HR@20 & 0.3139 & 0.3143 & 0.2940 \\
Test HR@20 & 0.1883 & 0.1878 & 0.1823 \\
Valid N@10 & 0.1878 & 0.1897 & 0.1647 \\
Test N@10 & 0.1139 & 0.1139 & 0.1006 \\
Valid N@20 & 0.1989 & 0.2006 & 0.1766 \\
Test N@20 & 0.1204 & 0.1208 & 0.1086 \\
\hline
\end{tabular}
\label{tab:baby_smaller_omega}
\end{table}

\begin{table}[htbp]
\caption{Average text–text and image–image similarity for Baby, Electronics, and Sports datasets.}
\centering
\begin{tabular}{lcc}
\hline
\textbf{Dataset} & \textbf{Text–Text Similarity} & \textbf{Image–Image Similarity} \\
\hline
\(Baby\)         & 0.2627 & 0.2239 \\
\(Electronics\)  & 0.1867 & 0.2348 \\
\(Sports\)       & 0.2085 & 0.2183 \\
\hline
\end{tabular}
\label{tab:similarity_results}
\end{table}

On the larger \(Elec\) dataset, adding user embeddings leads to a lower performance than the original MuSTRec system. Again this is likely due to the greatly enhanced data volume compensating for issues of interaction sparsity.
We considered the possibility that perhaps different datasets had different characteristics that made them more or less amenable to this approach. For example, datasets with many repeated or very similar items. 
However, we determined that none of the datasets have repetitions between the training, validation or test sets for any user. In Table \ref{tab:similarity_results}, we also calculated the item feature cosine similarity for all three datasets, but these do not indicate a significant difference. 
We explore this further by downsampling the \(Elec\) dataset randomly to the size of \(Baby\). As can be observed in Table \ref{tab:downsampled_elec} this led to once again observing huge performance gains from the introduction of user embeddings.  This confirms that the effect stems from the dataset size and interaction scarcity.

\begin{table}[htbp]
\caption{Downsampled Elec Results}
\centering
\small
\begin{tabular}{lcccc}
\hline
\textbf{Metric}  & \textbf{\(\omega=0.001\)} & \textbf{\(\omega=0.01\)} & \textbf{\(\omega=0.1\)} & \textbf{\(\omega=1\)} \\
\hline
Valid HR@10 & 0.2046 & 0.1611 & 0.0471 & 0.0197 \\
Test HR@10 & 0.1030 & 0.0822 & 0.0276 & 0.0131 \\
Valid HR@20 & 0.2236 & 0.1868 & 0.0623 & 0.0332 \\
Test HR@20 & 0.1144 & 0.1004 & 0.0372 & 0.0269 \\
Valid N@10 & 0.1621 & 0.1171 & 0.0301 & 0.0105 \\
Test N@10 & 0.0807 & 0.0588 & 0.0169 & 0.0063 \\
Valid N@20 & 0.1669 & 0.1236 & 0.0339 & 0.0139 \\
Test N@20 & 0.0836 & 0.0634 & 0.0193 & 0.0098 \\
\hline
\end{tabular}
\label{tab:downsampled_elec}
\end{table}

Finally, Table \ref{tab:baby_nomulti} shows the result of disabling the multimodal component losses while still passing the user embeddings into the sequence. 
Under this regime, the user-item and item-item graphs are not updated, and the user embedding is only weakly updated through its integration with the sequential objective. Surprisingly, we see a complete collapse in the value of the user embeddings here.
This further suggests that the boost in performance was not simply due to letting the transformer learn user-specific embeddings, but rather due to the complex interaction of the GNN and the collaborative filtering signal across related users.

\begin{table}[htbp]
\caption{Results on the Baby dataset for MuSTRec and MuSTRec-U, both with the Multimodal Component Loss disabled}
\centering
\small
\begin{tabular}{lcccc}
\hline
\textbf{Metric}  & \textbf{MuSTRec-U} & \textbf{MuSTRec} \\
\hline
Valid HR@10 & 0.0497 & 0.0629 \\
Test HR@10 & 0.0353 & 0.0489 \\
Valid HR@20 & 0.0734 & 0.0960 \\
Test HR@20 & 0.0547 & 0.0728 \\
Valid N@10 & 0.0273 & 0.0355 \\
Test N@10 & 0.0194 & 0.0275 \\
Valid N@20 & 0.0333 & 0.0439 \\
Test N@20 & 0.0243 & 0.0335 \\
\hline
\end{tabular}
\label{tab:baby_nomulti}
\end{table}

\section{Conclusion}
In this paper, we presented MuSTRec, an approach to unifying user-item and multimodal item-item graphs with transformer-based sequential prediction heads. Our experiments showed that embeddings from multimodal graphs fused with frequency-based self-attention mechanism can substantially boost recommendation accuracy compared to previous state-of-the-art. Ablation studies confirmed the importance of the graph embeddings. Integrating user embeddings was investigated and it was shown to yield substantial gains ($>200\%$) on smaller datasets.

\appendix
\clearpage
\section{Additional Ablation Figures}

\begin{figure}[hbp]
    \centering
    \includegraphics[width=0.95\columnwidth]{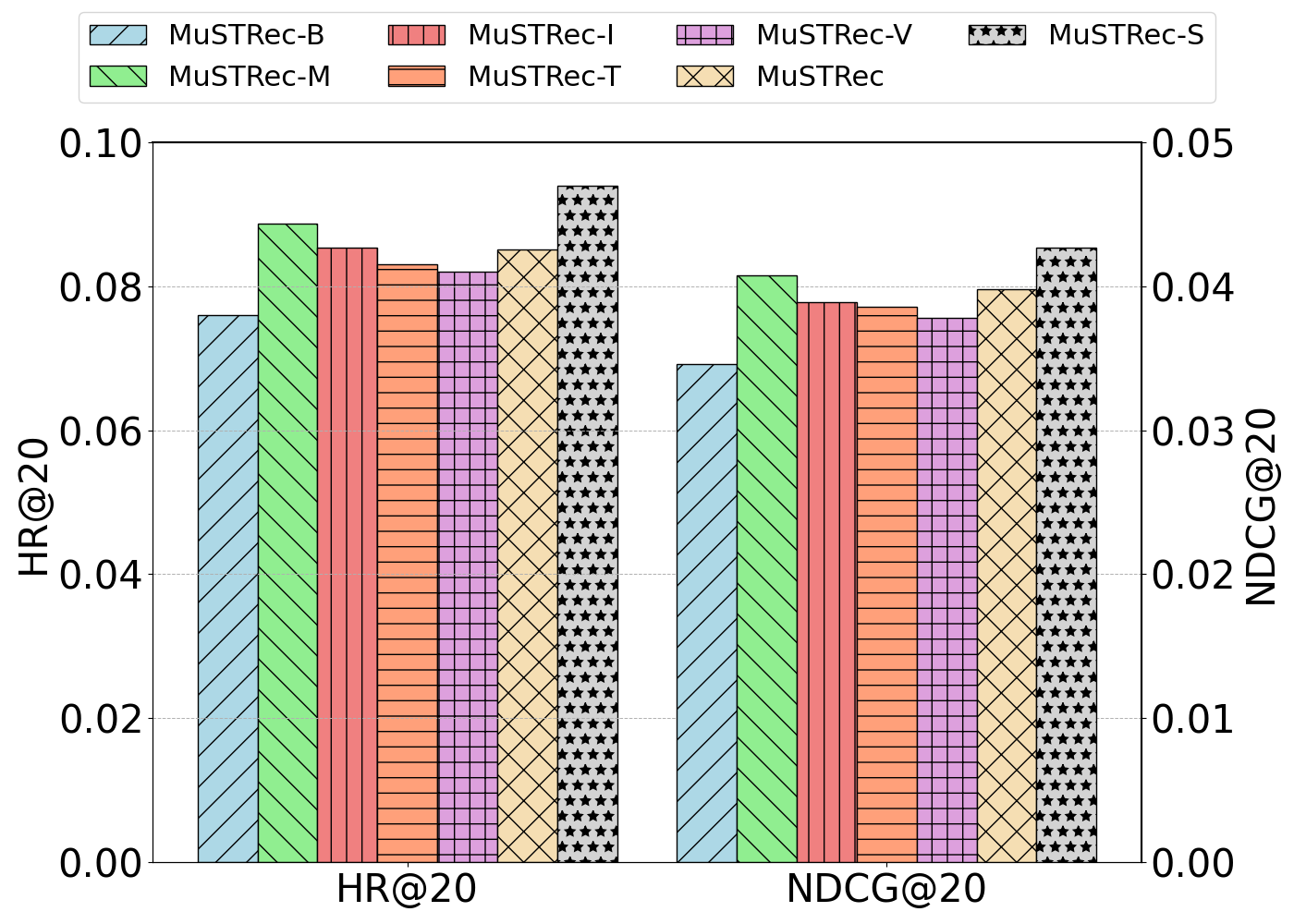}
    \caption{\(Baby\) Ablation
    }
    \label{fig:baby_ablation}
\end{figure}

\begin{figure}[hbp]
    \centering
    \includegraphics[width=0.95\columnwidth]{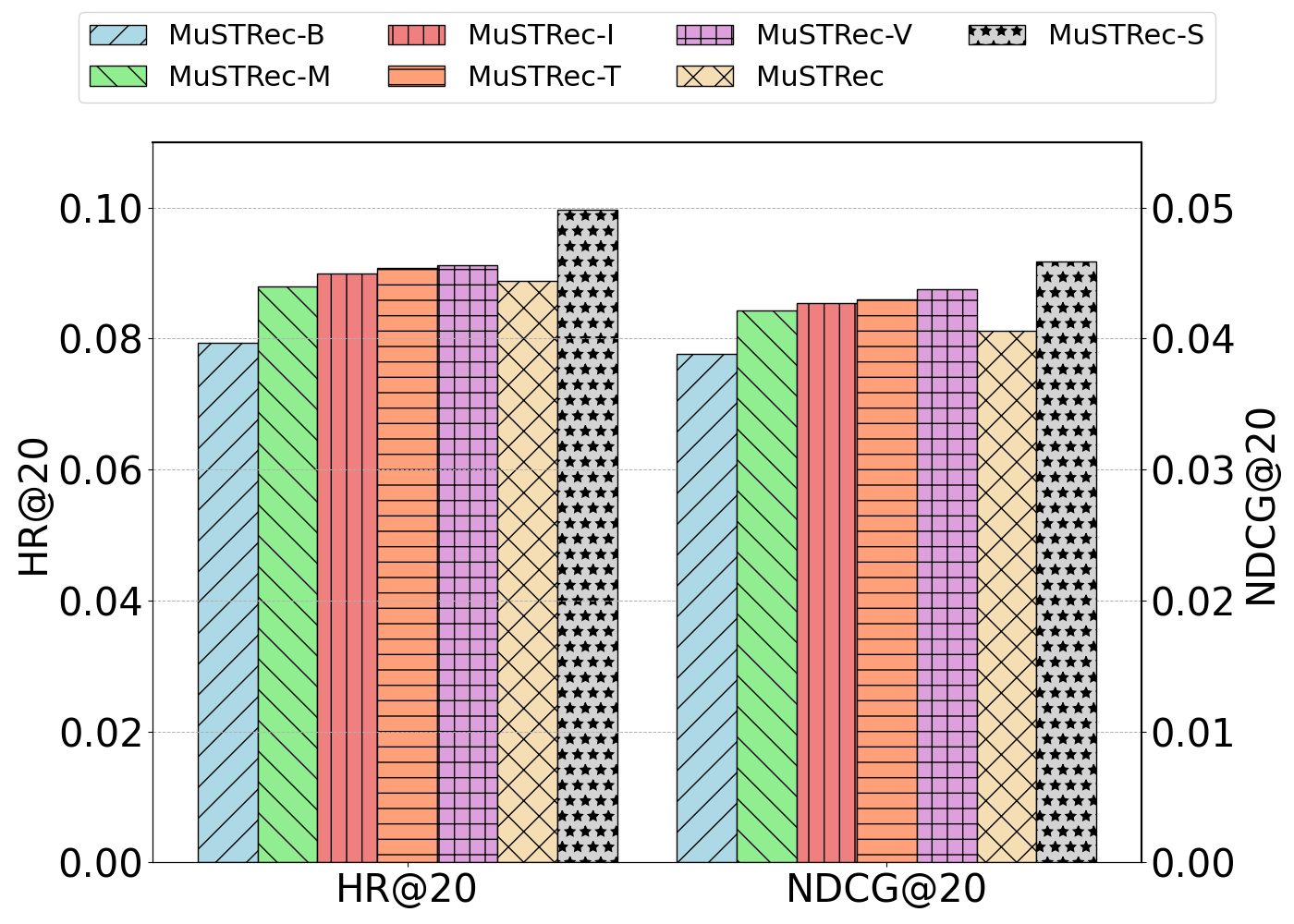}
    \caption{\(Sports\) Ablation
    }
    \label{fig:sports_ablation}
\end{figure}

\begin{figure}[!hbp]
    \centering
    \includegraphics[width=0.95\columnwidth]{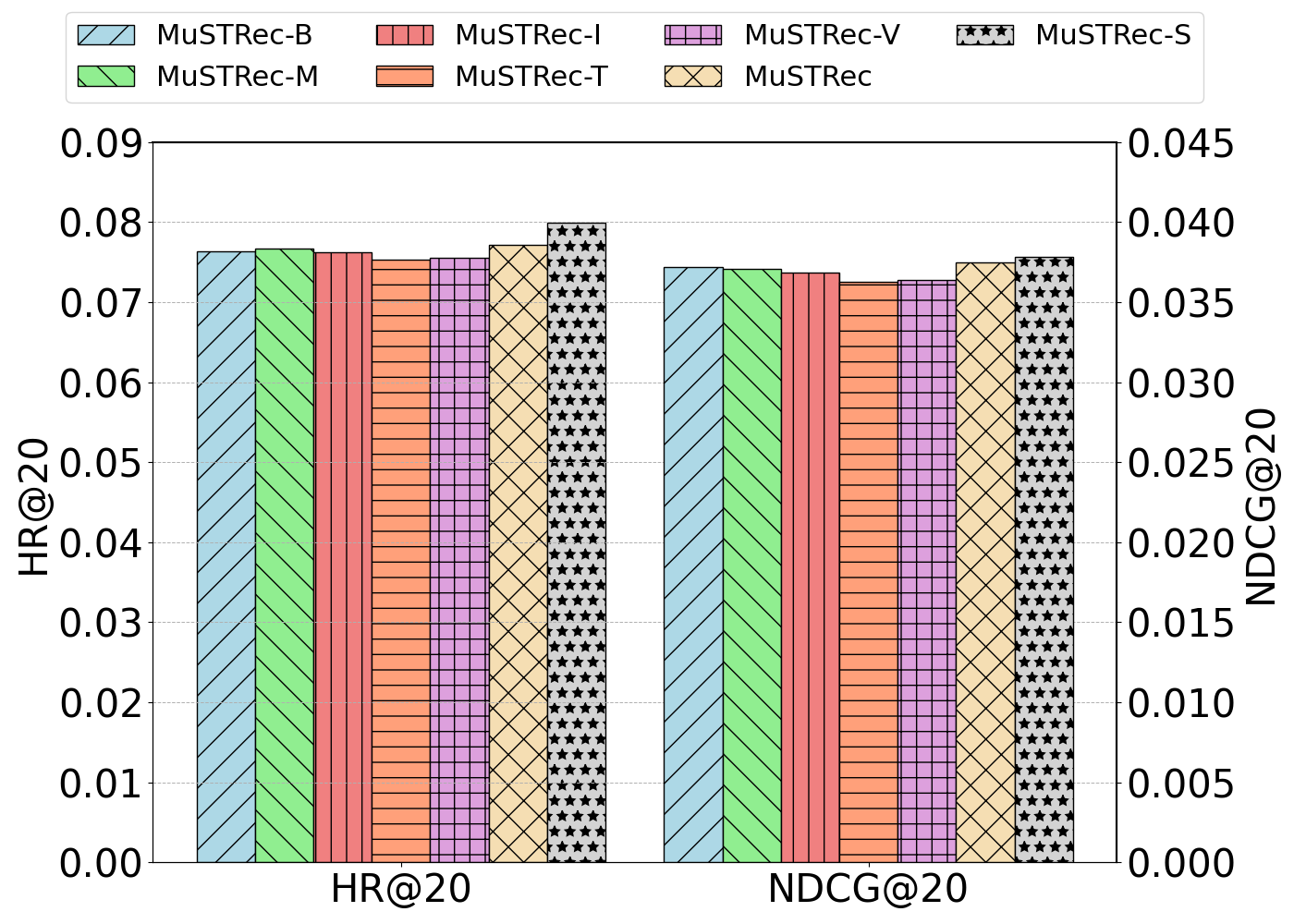}
    \caption{\(Elec\) Ablation
    }
    \label{fig:elec_ablation}
\end{figure}

Figures \ref{fig:baby_ablation}, \ref{fig:sports_ablation} and \ref{fig:elec_ablation} show ablation study results on \(Sports\), \(Clothing\) and \(Elec\) datasets.

\clearpage

\bibliographystyle{ACM-Reference-Format}
\bibliography{samples/sample-base}

\end{document}